\documentclass[12pt]{elsarticle}
\def\la{\langle}
\def\ra{\rangle}

\def\be{\begin{equation}}
\def\ee{\end{equation}}

\usepackage{graphics,graphicx}
\journal{Physica A}

\begin{document}

\title
{Unusual scaling in  a discrete quantum walk with random long range steps}

\author{Parongama Sen}
\address{Department of Physics, University of Calcutta, 92 Acharya Prafulla Chandra Road, Kolkata 700009, India.}

\begin{abstract}

A discrete time quantum walker is considered in one dimension, where at each step, the 
translation can be more than one unit length chosen randomly. 
In the simplest case, the probability that the distance travelled  is $\ell$ is
taken as  $P(\ell) = \alpha \delta(\ell-1) + (1-\alpha) \delta (\ell-2^n)$ with $n \geq  1$.  
Even the $n=1$ case shows a drastic change in the scaling behaviour
 for any $\alpha \neq 0,1$. Specifically,   $\la x^2\ra
 \propto t^{3/2}$ for $0  < \alpha < 1$, implying  the walk is slower compared to  the usual quantum walk.
This scaling behaviour, which is neither conventional quantum nor classical,  can be justified using a simple form for the probability density.  
The decoherence  effect is  characterized by two parameters
which vanish in a  power law manner close to $\alpha =0$ and $1$ with an exponent $\approx 0.5$.  
It is also shown that randomness is the essential ingredient for 
the decoherence effect. 

\end{abstract}

\maketitle

\section{Introduction}

Discrete time quantum  walk (DTQW)  is a phenomenon in which a random walker has 
a coin (also called chiral) degree of freedom  which dictates the translational 
motion of the walker at each discrete time step \cite{Feynman,Aharonov,kempe,nayak,amban,Venegas}. In contrast to the classical case, the walk may propagate
in different directions simultaneously as the walker may exist in a superposition of  coin states. 
 The time dependence of  the square of the displacement for a quantum walk is $\la x^2 \ra \propto t^2$  showing it is  much faster than the classical walker (where $\la x^2 \ra \propto t$), and hence can play a key role in  many
dynamical  processes. 
Apart from the discrete walk, the continuous time quantum walk has also been conceived \cite{farhi} where the coin degree of freedom is not present.  The 
speeding up over classical walk is noted in both  discrete and continuous walks.  

The  quantum walk can be  slowed down by decoherence effects 
\cite{kendon};
 in most cases this leads to a transition to a classical diffusive walk. 
Decoherence and subsequent localisation can take place   due to several factors like  
randomness in the environment,  
defects in the embedding lattice or graphs,
  measurements  of the 
position or chirality of the walk, inertia,  etc. Decoherence  has been studied  in one dimensional  
\cite{kendon1,brun1,brun2,shapira,romanelli1,proko,ermann,romanelli2,abala,romanelli3,annabestani,romanelli4,xiong2-1d,chenSR} and  two dimensional 
 \cite{chenSR,kosik2d,oliveia2d,gonulol2d,chandra2d,chen2d}
discrete walks as well as in 
 the continuous quantum walk  \cite{keating,Yueyin,salimi}. 
Such noisy quantum walks can even be non-unitary \cite{kendon,xiong}. 
Experimentally, cases with both static and dynamic disorder have been studied \cite{Schreiber1}  resulting  
in   Anderson localization type phenomena and diffusive behaviour respectively. 
In  the cases studied so far, the results are strongly dependent on the  parameters controlling   the decoherence. 

In most of the earlier works in one dimension, the disorder has been 
incorporated through the coin operator 
in different ways. In certain cases,   stochasticity  in position space has also been 
considered, for example, by breaking links. 
In this work, we consider a discrete time quantum walk on  a line with 
dynamic disorder  in the  translational motion.
%

Usually, it is assumed that the translation of the quantum walker is
of equal length at each time step.
 We relax this condition by allowing the  translation through 
a distance  $\ell \geq 1$, chosen randomly, at each time step. 
Such long range hopping has been considered in quantum transport phenomena by 
including interaction with a thermal bath of 
oscillators 
 in a tight binding model 
 \cite{caceres} to study decoherence effects.  
While the longer steps can make the transport faster, the randomness will also
have its effect by slowing it down. In this paper, we intend to investigate the result of this competition in the quantum walk on a line.
The probability distribution of the position of the quantum walker 
 and its moments are calculated and the primary objective in the 
present paper 
is to compare the behaviour of these quantities  with the conventional classical random and non-random quantum walks.  

In section 2 we describe the quantum walk considered in the paper and the 
details of the quantities calculated. The results are presented and analysed in section 3. A concluding  section is added in the end.

\section{The random long ranged walk}

In the quantum walk in one dimension, the state
of the walker is expressed in the $|x\rangle \otimes |d\rangle$ basis, where $|x\rangle$
is the position (in real space) eigenstate and $|d\rangle$ is the
chirality  eigenstate (either left ($|L\ra$)  or right ($|R\ra$)). 
The state   of the particle, $\psi({x},t)$ can be written  as
\begin{equation}
\psi({x},t)=
\left[ {\begin{array}{cc}
    \psi_{L}({ x},t)\\
    \psi_{R}({ x},t) 
\end{array}} \right]
\end{equation}
\noindent
For the rotation in the chiral space,    we have used the 
Hadamard coin \cite{nayak,amban} unitary operator $H$  
represented  by 
\begin{equation}
H =\frac{1}{\sqrt{2}}
\left [ {\begin{array}{cc}
  1 & 1 \\
  1 & -1 
\end{array}} \right]
\end{equation}
The occupation probability of site $x$ at time $t$ is given by $f(x,t) =
|\psi_{L}({ x},t)|^2 + |\psi_{R}({ x},t)|^2$
\noindent
; sum of these probabilities over all $x$ is $1$ at each time step. 
 The walk is initialized at the origin with
$\psi_{L}(0,0) = a_0,  \psi_{R}(0,0)= b_0; ~~a_0^2 + b_0^2 =1$
and
$\psi_{L}(x\neq 0,t=0) =   \psi_{R}(x\neq 0,t=0)= 0$.


As long as the displacement  $\ell$ at each step is a constant, results are independent of $\ell$ apart from a trivial scaling factor. 
In the present case, the value of $\ell$ is chosen randomly 
from a distribution. 
The conditional translation operator at any time $t$ can then be written in a compact form:
\begin{equation}
T(t)=|R\ra \la R | \otimes \sum_x |x+\ell(t)  \ra \la x|+|L\ra \la L | \otimes \sum_x |x-\ell(t)  \ra \la x|.
\nonumber
\end{equation}

The distribution of $\ell$ can be chosen in many ways, we choose one which introduces longer step lengths in the simplest  possible manner. 
We  consider
only two possible step lengths: $\ell = 1$ and $\ell = l_{max} = 2^n$ with $n \geq  1$ ($n$ having a fixed value) and   
\begin{equation}
P(\ell) = \alpha \delta(\ell-1) + (1-\alpha) \delta (\ell-2^n).
\label{eq:prob}
\end{equation}

The limits $\alpha =1$ and $\alpha =0$ are equivalent to usual quantum walks and
it is sufficient to  consider the interval $0.5 \leq  \alpha <  1.0$  due to symmetry; $\alpha = 0.5$ is the case with maximum randomness and $\alpha =1$ corresponds to zero randomness.  
Note that in  a classical walk one can also introduce such a variation, for finite values of $n$, the scaling of the moments will remain the same due to central limit theorem. 

The walk is initialised with $a_0 = \sqrt{\frac{1}{3}}$ and $b_0 = \sqrt{\frac{2}{3}}$ 
such that in absence of disorder, an asymmetric probability density profile is obtained and one can study   the  scaling of both the first and second moments.
The results are averaged over 1000 configurations.

We have evaluated $f(x,t)$  for the long ranged walk and studied the scaling behaviour of the first two moments. In the present work, only one parameter controlling the randomness has been used.

\section{Results and analysis}

\subsection{Results for $n=1$}

\begin{figure}
\begin{center}
\includegraphics[width=7cm]{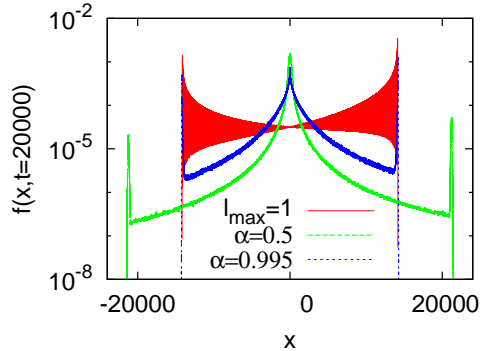}
\caption{$f(x,t)$ for  the quantum long ranged walk  with step sizes 1 and 2 with probability $\alpha$ and $ (1-\alpha)$ respectively. Results for $\alpha = 0.995$ and $0.5$ are compared with the case  $\alpha =1$ where the maximum step size $l_{max} = 1$. }
\label{fig1:comp}
\end{center}
\end{figure}

We first consider the case where step lengths are either 1 or 2 chosen according to eq.  (\ref{eq:prob}).  We note that as soon as $\alpha$ deviates from unity, the shape of $f(x,t)$ 
assumes a completely different form. $f(x,t)$ shows two ballistic peaks and an additional peak around zero, the latter is absent for the quantum walker. 
It may be mentioned that this shape is not a result of  averaging over different realisations,
even a single configuration shows these features. 
Fig. \ref{fig1:comp} shows the distributions for  $n=1$ with 
$\alpha = 0.5$ and $\alpha = 0.995$. 
The result  for $\alpha =1$ (i.e., $l_{max} =1$, the usual case),  is also shown for comparison. 
We also note that while  asymmetry is maintained in the ballistic peaks, 
the centrally peaked part is symmetric. 
As $\alpha$ is increased from 0.5, the width of the distribution 
decreases as  more steps with $\ell =1$ is taken compared to $\ell=2$. Although the range is almost same for $\alpha =1$ and $\alpha =0.995$, even the small deviation of $\alpha$ from unity is effective in making the distribution strikingly different from that at $\alpha =1$. 
As $\alpha$ approaches   $1$, the peak value of $f(x,t)$ at $x=0$    becomes 
less compared to the ballistic peaks in height. 

\begin{figure}
\begin{center}
\includegraphics[angle=0,width=7cm]{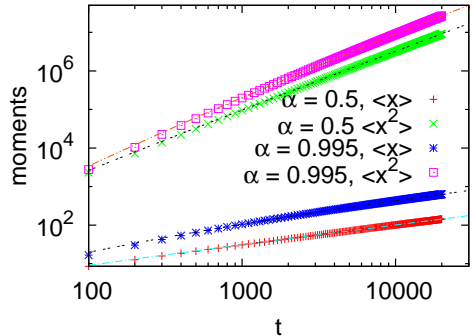}
\caption{The first two moments of $f(x,t)$ for two extreme values of $\alpha$ are shown ($n =1$). The continuous lines are best fit curves obtained using the  the form  given in  Equations \ref{eq:mom1} and \ref{eq:mom2}.
}
\label{fig2:fits}
\end{center}
\end{figure}

Plotting the first two moments, $\la x\ra$ and $\la x^2 \ra$, we note a drastic change in the scaling forms compared to $\alpha =1$. For any value of $\alpha \neq 1$, the asymptotic behaviour suggests $\la x\ra \propto t^{1/2}$ and 
$\la x^2 \ra \propto t^{3/2}$ shown in Fig. \ref{fig2:fits}. 
The scaling of the fluctuations $\la x^2 \ra - \la x\ra ^2$ follows the 
same scaling form as $\la x^2\ra$. 
This indicates   that  the quantum walker, when allowed to take long range steps randomly, ends up being slower than usual.

\begin{figure}
\begin{center}
\includegraphics[width=7cm]{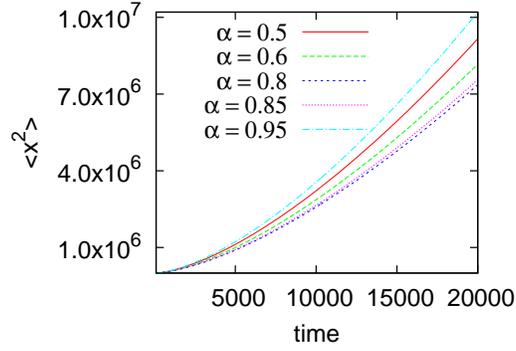}
\caption{The second moment of the distribution $f(x,t)$ shown against time 
for different values of $\alpha$ (n=1). It shows a minimum value at $\alpha \simeq 0.8$. }
\label{fig3:diff}
\end{center}
\end{figure}

We also note that the  two ballistic peaks of  $f(x,t)$ 
occur approximately  at $x = \pm (2-\alpha)t/\sqrt{2}$ 
which corresponds to a simple weighted linear superposition of the two cases for $\ell = 1$ and $\ell = 2$ for the usual DTQW. 
Hence the range decreases linearly with $\alpha$ which is to be expected. However, although $\la x^2 \ra \propto t^{3/2}$ for all $\alpha < 1$, the actual value of $\langle x^2 \rangle$  
varies non-monotonically with $\alpha$ (see Fig. \ref{fig3:diff}); 
as $\alpha$ is increased from 0.5, it first shows a  decrease but then increases  beyond $\alpha \approx 0.8$.
This indicates the walker is localised maximally for a value of $\alpha \simeq 0.8$ and not at $0.5$ as one would naively expect. 
We will get back to this point later.

\begin{figure}
\begin{center}
\includegraphics[width=5.2cm]{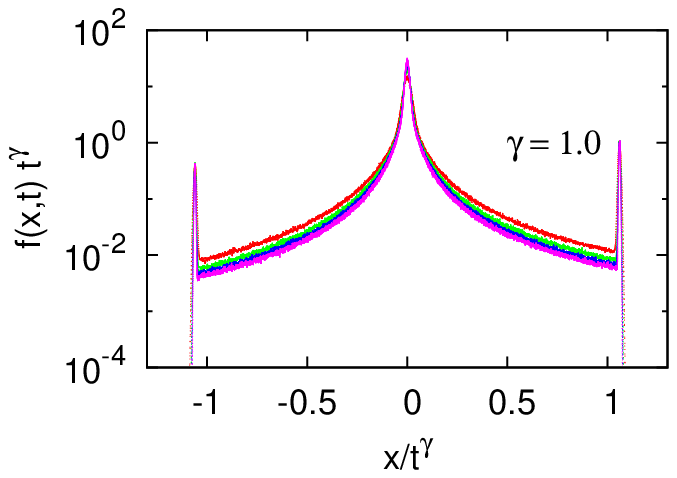}
\includegraphics[width=5.2cm]{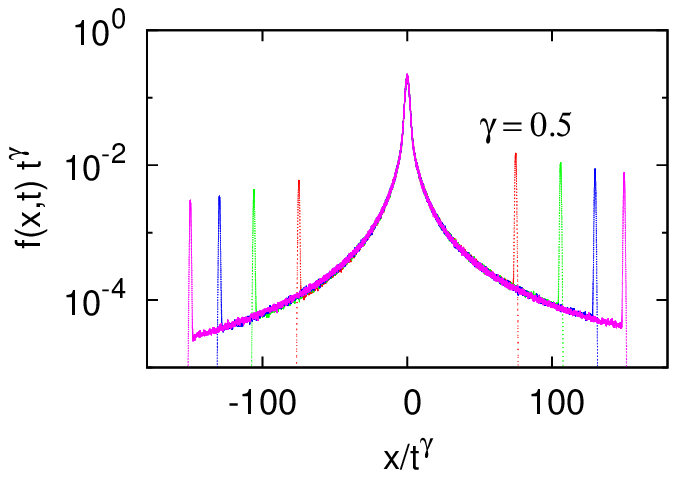}

\includegraphics[width=5.2cm]{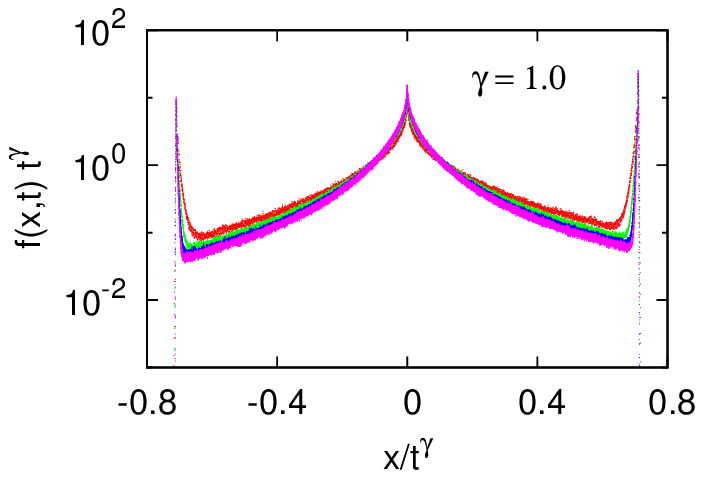}
\includegraphics[width=5.2cm]{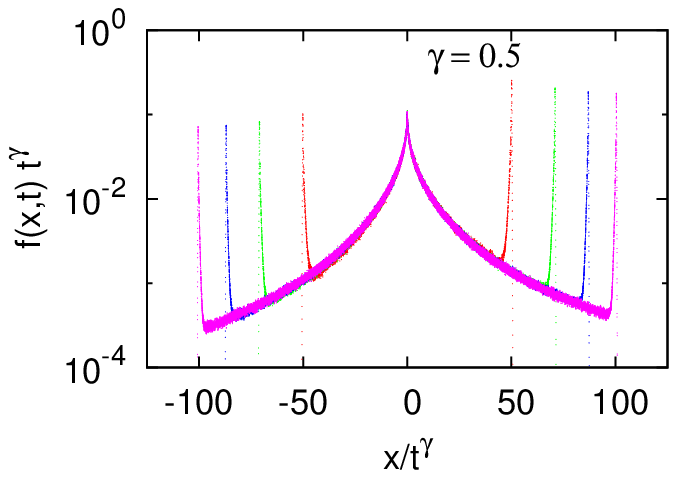}
\caption{Data collapse of $f(x,t)$ for four different values of $t$ for the  ballistic  peaks and the central peak for $\alpha = 0.5$ (top panel) and $\alpha = 0.995$ (bottom panel). All data are for $n=1$.}
\label{fig4:scaling_cent}
\end{center}
\end{figure}

Analysing the probability densities,  one finds that $f(x,t)$ shows two distinct scaling behaviour: plotting
$t^\gamma f(x,t)$ against the scaled variable $x/t^\gamma$,  
the central part collapses with $\gamma = 0.5$, while with $\gamma=1$ the extreme values show a collapse;
this happens for all values of $\alpha \neq 1$; data for 
two extreme values are   shown in Fig. \ref{fig4:scaling_cent}.

In order to obtain an approximate estimate of the moments, the problem one faces is that no     
 simple form of the distributions  exists. Still, one can  make a rather gross  approximation by taking  
  $f(x,t)$  (for $\alpha \neq 1$)   to be a discrete 
function with non-zero values at three points only: $x=0$ (where the central peak occurs) and $x = \pm ct$ 
(where the ballistic peaks occur). Noting the peak values  scale as $1/\sqrt{t}$ and $1/t$ at these three points, one can write $f(x,t)$ as 

\begin{equation}
f(x,t) = a_1 \frac{1}{t}\delta(x-ct) + 
 a_2 \frac{1}{t}\delta(x+ct) + a_3\frac{1}{\sqrt{t}}\delta(x).
\label{eq:dist}
\end{equation}
In general,   $a_1 \neq a_2$ as  the walk is asymmetric. 
From Eq. (\ref{eq:dist}), one can estimate the first two moments as 

\begin{equation}
\langle x\rangle = t/(b_1+{b_2}\sqrt{t})
\label{eq:mom1}
\end{equation}
and
\begin{equation}
\langle x^2\rangle = t^2/(b_3+b_4\sqrt{t}),
\label{eq:mom2}
\end{equation}
where $b_i$ are related to the constants $a_i$ and $c$. 
%

The above forms of $\langle x\rangle$  and $ \langle x^2 \rangle$ are  consistent with their  asymptotic bahaviour obtained numerically.
 More importantly, 
fitting the moments using equations \ref{eq:mom1} and \ref{eq:mom2},  we find excellent agreement over the entire time range with errors in the estimates less than one percent in general,
shown in Fig. \ref{fig2:fits}.

\begin{figure}
\includegraphics[angle=0,width=7cm]{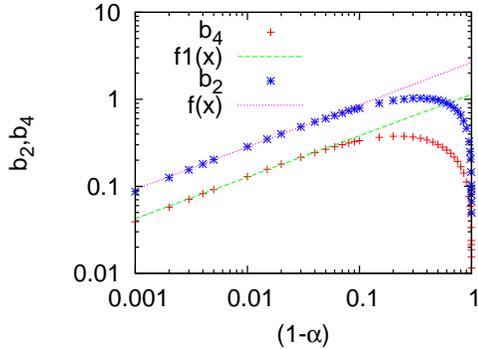}
\caption{The  decoherence parameters vanish following a power law close to   $\alpha=1$. Here $f(x) = 2.66(1-\alpha)^{0.487}$ and $f1(x) = 1.14(1-\alpha)^{0.476}$. These data are for $n=1$. }
\label{fig5:decoh}
\end{figure}

Mathematically, assuming Eq. (\ref{eq:dist}) is correct,  the change in behaviour in the scaling compared to the DTQW is obviously due to the existence of the central peak. The  latter  results in the terms 
proportional to $\sqrt{t}$ with  coefficients  $b_2$ and 
$b_4$ in  the denominators of equations (\ref{eq:mom1})  and (\ref{eq:mom2}),  in the absence of which one would recover the usual behaviour of the DTQW. 
For $\alpha =1$, both $b_2$ and $b_4$ should vanish.
We identify the parameters $b_2$ and $b_4$ as decoherence parameters
and study their behaviour with $\alpha$. 
Attempting a form  $ (1-\alpha)^\beta$,   
 a good agreement with $\beta \approx 0.5$ is found for both $b_2$ and $b_4$ 
shown in Fig. \ref{fig5:decoh}. 
Of course, symmetry demands that  $b_2, b_4$ also vanish at  $\alpha =0$ 
with the same exponent, this  has been   been verified.  Clearly, 
there will be a peak value occurring for $b_2$ and $b_4$ for $0 \leq \alpha \leq 1$.    
 The peak  values   do not occur at $\alpha = 0.5$; for $b_4$, the maximum is at $\alpha \approx 0.8$. This is consistent with the  observation that
at this point   $\langle x^2 \rangle$ has the minimum value. 
 For $b_2$, the peak value is close to $\alpha = 0.65$. 

\begin{figure}
\includegraphics[angle=0,width=7cm]{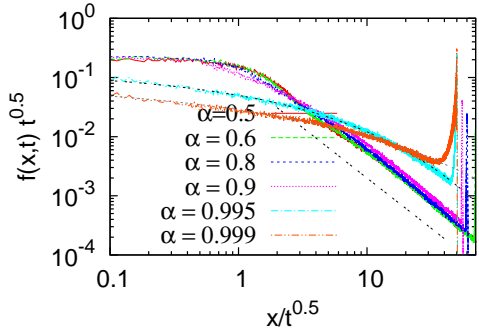}
\caption{Behaviour of $f(x,t)$ shown for $x > 0$ ($n=1$). The power law decay occurs for $x/\sqrt{t} > 1$ over a finite region. The dashed line has slope equal 
to $-1.7$. The curves for $\alpha = 0.995$ and $0.999$ are fitted to the functions
$0.13\exp(-0.97x^{0.39})$ and $0.08\exp(-1.07x^{0.30})$ respectively. }
\label{fig6:zero}
\end{figure}

 As the centrally peaked region, assumed to behave as a delta function, is the key to
the decoherence, we probe further into the region close $x=0$. Usually, 
  decoherence effects show that the emerging centrally peaked part tends to a Gaussian form 
with time such that a classical behaviour (i.e. $\la x^2\ra \propto t$) is obtained at large times. 
(For cases where Anderson localisation takes place, the behaviour is exponential \cite{Schreiber1}.)
In this case, however, we find that the behaviour close to $x=0$ is definitely non-Gaussian. 
Rather, it shows a behaviour compatible with a power law decay for  
$\alpha \leq 0.9$ 
for 
$x/\sqrt{t} > 1$ while  for smaller values of $x/\sqrt{t}$ it is almost a constant. The curves also show negligible dependence on $\alpha$. 
However, the power law is valid only up to a finite value of  $x/\sqrt{t}$, which   depends on $\alpha$.       
For  $\alpha$ values closer to unity, we observe that  a slower stretched exponential decay fits  a wider region of  $x/\sqrt{t}$. These results are shown 
in Fig.  \ref{fig6:zero}.   The change in behaviour from power law to stretched exponential may be an effect of the closeness to $\alpha =1$ point, where transition to a pure quantum walk occurs.  
The above observation shows that although  the decay is power law for most values of $\alpha$, as the behaviour is limited to a finite region, and the associated exponent ($\simeq 1.7$) is not too small, the assumption that it is a delta function has worked well. 

\subsection{Results for $n > 1$}

One difficulty is  as $n$ increases, it is not possible to
continue for very large times in the numerical simulations due to memory limitations.
We have therefore considered  up to $n=3$.  
For values of $n = 2$ and $3$,  $f(x,t)$ shows the existence of  more peaks at intermediate points along with the two ballistic peaks and the central peak for all $ 0.5 \leq \alpha < 1$.
 The secondary peaks are however not  as sharp as those obtained for $n=1$. Apparently,  the sharpness decreases
systematically with $n$.
The number of secondary peaks on either sides also increases with $n$, for $n=1,2,3$ the number of prominent peaks
appears to be equal to $n$ on each side.

\begin{figure}
\includegraphics[width=5cm]{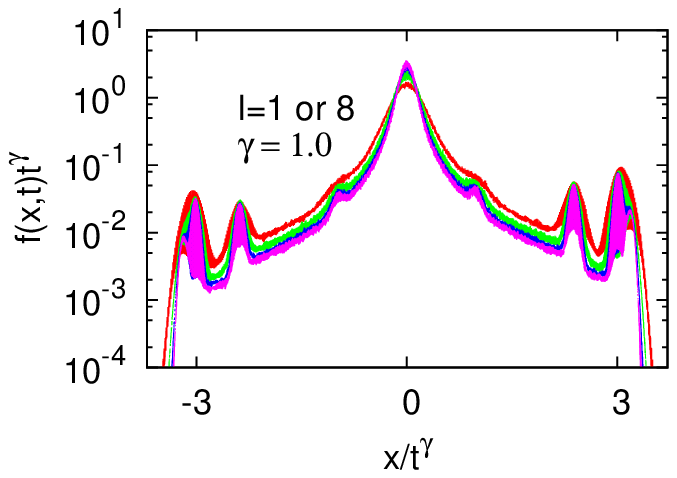}
\includegraphics[width=5cm]{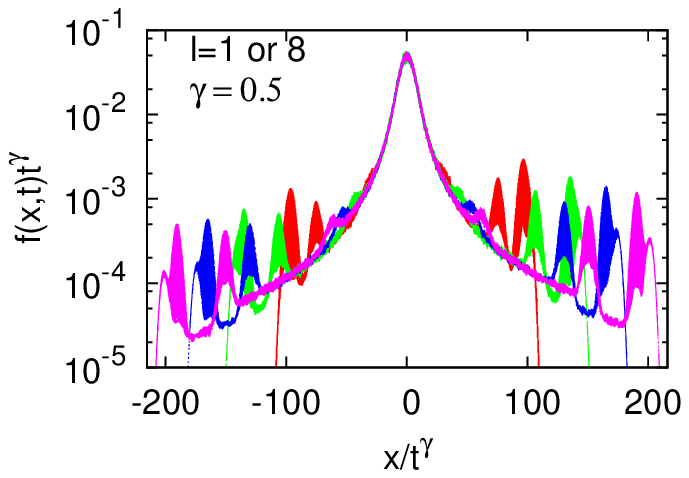}

\includegraphics[width=5cm]{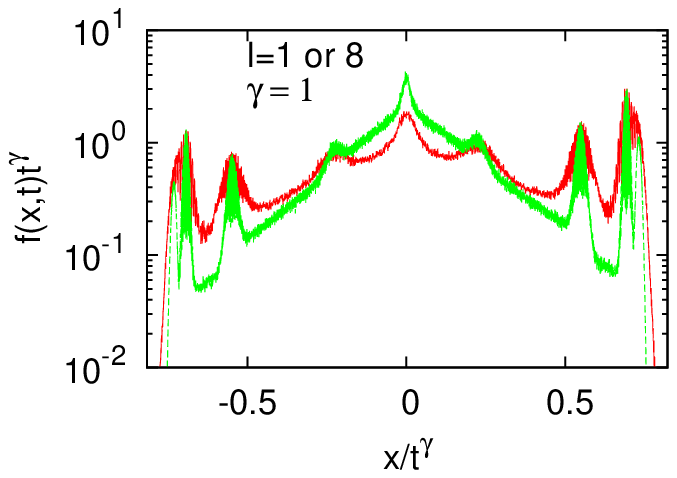}
\includegraphics[width=5cm]{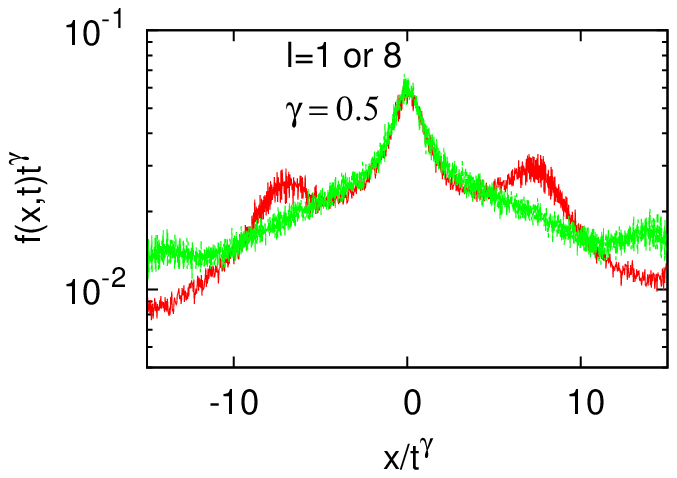}
\caption{Scaled $f(x,t)$ for  the quantum long ranged walk  with step sizes 1 and 8 with probability $\alpha$ and $ (1-\alpha)$. Results for $\alpha = 0.5$ (top panel, with data for four $t$ values) and $0.995$ (bottom panel, with data for two $t$ values) are shown with two trial
values of the scaling parameter $\gamma$.}
\label{fig7:n1n8}
\end{figure}

As for $n=1$, here too we have attempted  
 scaling  collapses of  $f(x,t)$
at short and long ranges  (Fig. \ref{fig7:n1n8} shows some examples for $n=3$). 
While  the collapse of the centrally peaked region
is very good for $\alpha = 0.5$,  the collapse for the secondary  peaks 
are not not so accurate in comparison; 
only the tips of the peaks seem to collapse. 
The quality of the collapses  becomes poorer for $\alpha = 0.995$.  


\begin{figure}
\includegraphics[angle=0,width=6cm]{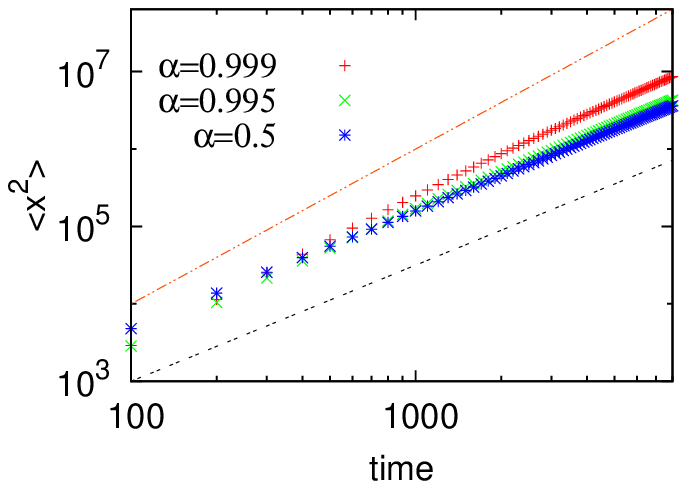}
\includegraphics[angle=0,width=6cm]{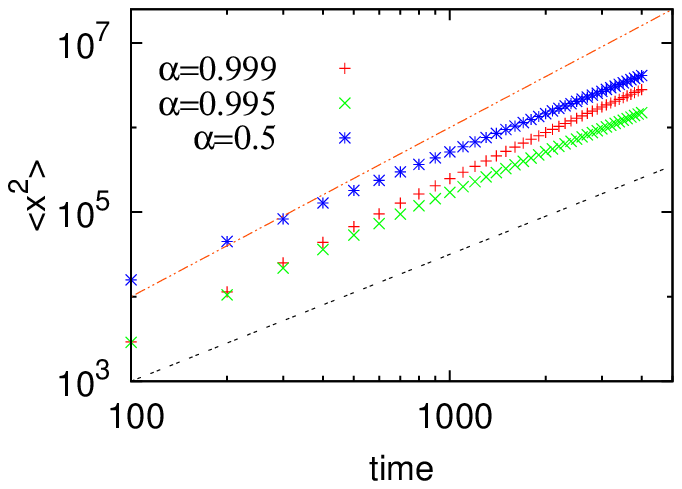}
\caption{The second   moment of $f(x,t)$ for $n=2$ (left) and $n=3$ (right) for several values of $\alpha$. The continuous lines have slopes 2 and 1.5.}.
\label{fig8:moms}
\end{figure}

Although the data collapses are not as good in quality as for $n = 1$, 
 $\la x^2 \ra$ for $n=2$ and $3$  (shown in Fig. \ref{fig8:moms})
again   scale as  
$t^{3/2}$ asymptotically.  For  $\alpha = 0.999$, which is very close to unity, the initial variation is like $t^2$ but it crosses over to a 
behaviour consistent with the exponent value equal to 1.5. This again shows that the
minimum randomness drives the system to a novel scaling behaviour.

\subsection{Long ranged  non-random walk}

In the usual quantum walk (without randomness), the interference effects are responsible for the non-classical behaviour. 
To understand the present  results, we argue that when the 
step lengths  are randomly chosen,  the  interference effect is suppressed. 
For example there will be  no interference  when a step of unit length is followed by a step of length 2. 
Another reason may be that as longer steps are taken, the walker, as it moves both ways, can find itself closer to the origin   
with higher probability even at larger times. 
However, a different walk, discussed in this subsection,  shows that this may be a necessary condition only and not sufficient.

\begin{figure}
\includegraphics[angle=270,width=6cm]{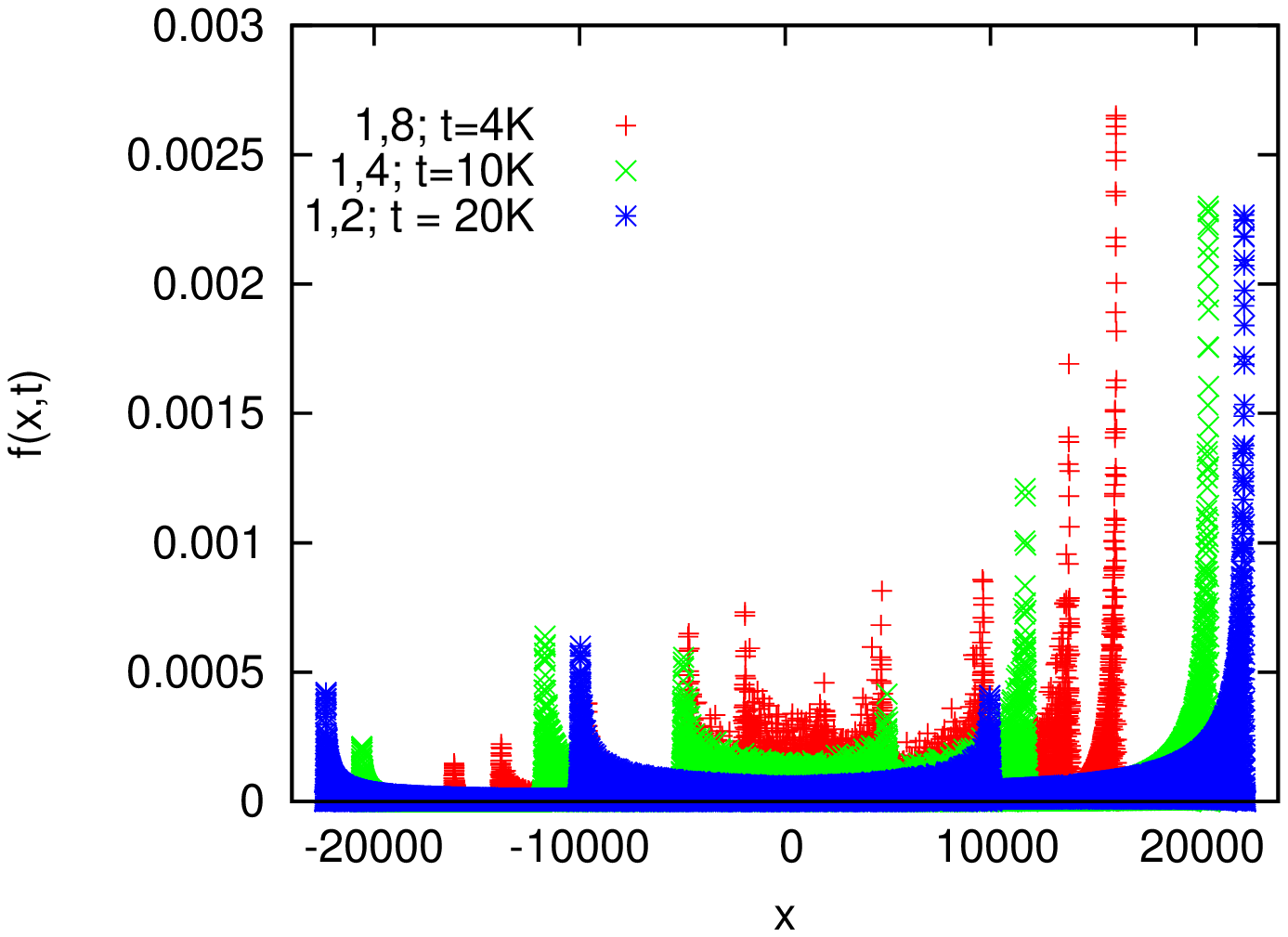}
\includegraphics[angle=270,width=6cm]{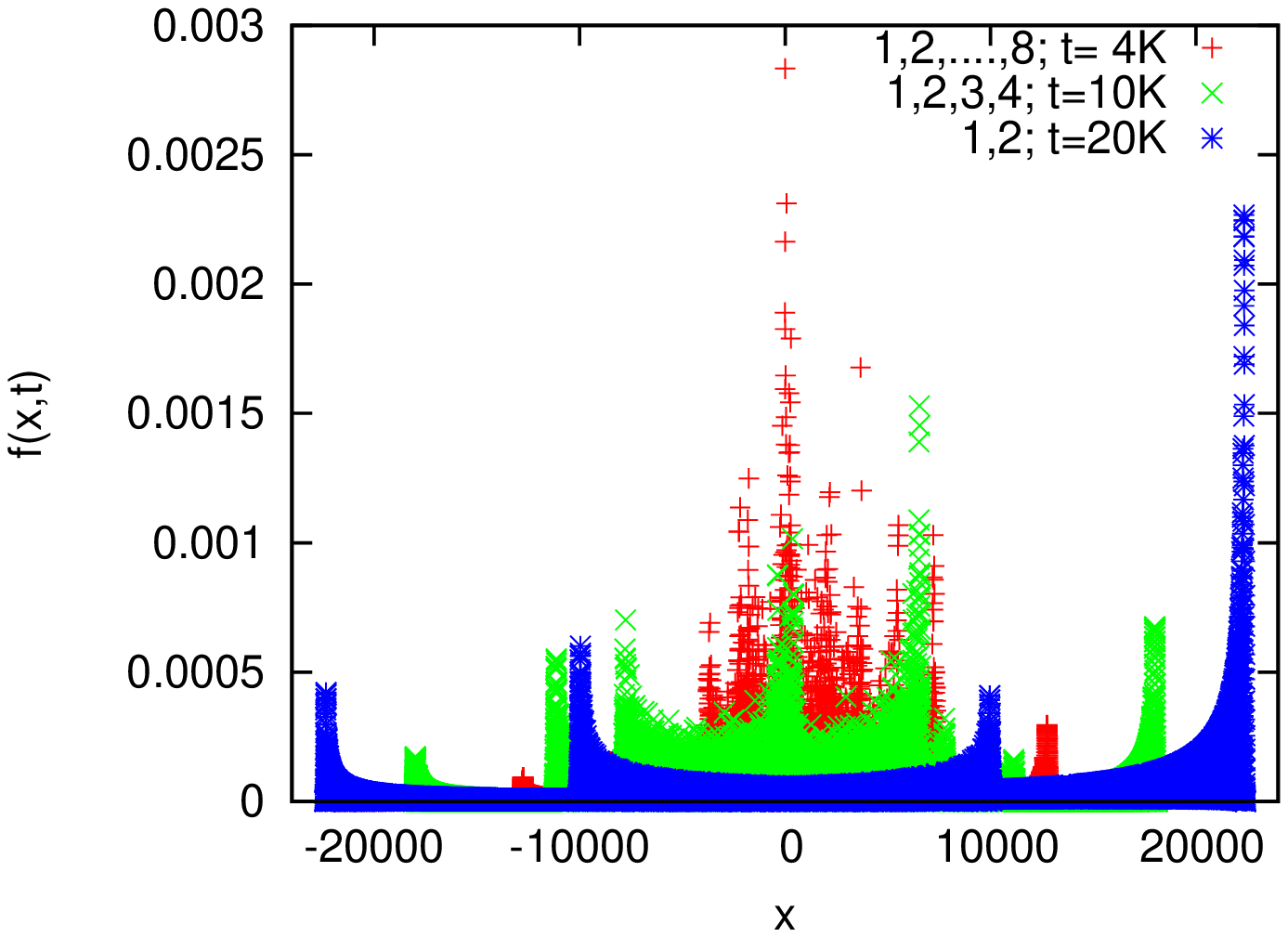}
\caption{
$f(x,t)$ for the long ranged walk without disorder shown for several cases. 
In the left panel, 
data for walks of periodicity 2 are shown while the data for periodicity $> 2$ 
are shown in the right panel . 
 The step lengths within one period are indicated in the key. The case for the walk 1,2,1,2.... is shown in both panels for a comparison.}
\label{fig9:benn-dist}
\end{figure}

\begin{figure}
\includegraphics[angle=0,width=6cm]{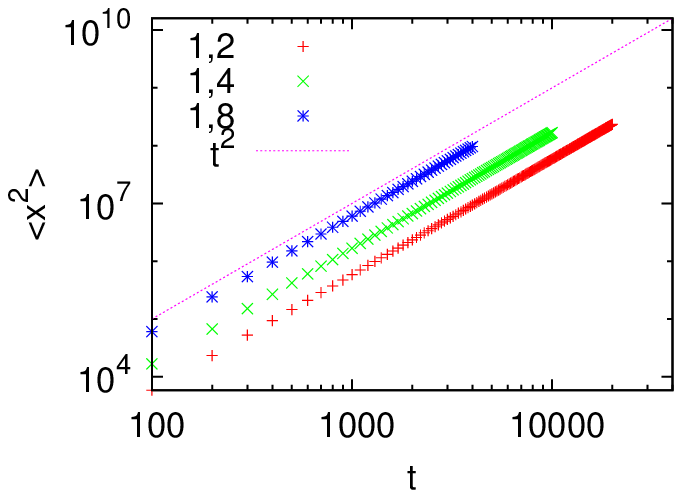}
\includegraphics[angle=0,width=6cm]{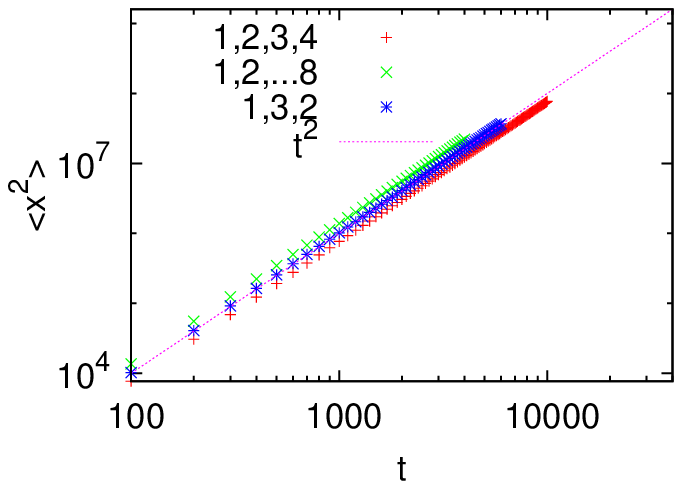}
\caption{Second 
moments calculated  for walks with long range steps but no randomness 
show  usual scaling $\la x^2 \ra \propto t^2$. Left panel - periodicity 2,
right panel periodicity $> 2$.}
\label{fig10:benn-moms}
\end{figure}

In this walk,  longer steps are allowed 
but there is no randomness. 
The steps here are 
taken in an ordered manner,   for example, the walk may comprise of steps $1, 2, 1, 2, \dots$. 
We note that this scheme does bring some changes  in the probability density, as secondary peaks emerge (Fig. \ref{fig9:benn-dist} shows some results for walks of different 
periodicities). 
However,   the second moment still shows the conventional 
scaling; $\langle x^2 \rangle \propto t^2$. 
The variation of $\la x^2 \ra$  with time $t$ is shown in Fig. \ref{fig10:benn-moms}.  For walks which are comparable to the $n=1,2$ and $3$ random cases, i.e., walks with 
periodicity 2 as steps alternate between 1 and $2^n$,  the results are shown in the left panel.   
For  walks of periodicity $>2$, the data plotted in the  
  right panel of Fig. \ref{fig10:benn-moms} show similar behaviour.

\section{Summary and conclusions}

A numerical study to investigate  the effect of  randomly chosen step lengths in a quantum walk  in one dimension has been presented in this paper. 
The  randomness introduced in the step length of the quantum walk 
drastically changes its nature. There is a decoherence effect but unlike in the
cases studied so far, this effect does not drive it 
to a simple diffusive walk. 

We find a new scaling behaviour $\la x^2 \ra \propto t^{3/2}$ in the asymptotic limit. It is also observed that the minimum amount of disorder can 
drive the system to this sub-ballistic but super-diffusive scaling.
 The form of the distribution for $\ell$ given by eq. (\ref{eq:prob}) ensures that  the walker at 
the farthest position from the origin will not suffer any interference and hence one can expect a non-diffusive behaviour. In fact the ballistic peaks 
of $f(x,t)$, arising  out of this dynamics show a collapse when $x$ is scaled by $t$. On the other hand,  
 a non-Gaussian peaked region close to  the origin, which collapses when $x$ is scaled by $t^{1/2}$, is also present.  The result of the two effects 
is the novel scaling behaviour. One can compare this with the case 
encountered in \cite{proko}, where a similar peak around the origin was noted, 
 but in that case, it was not strong enough to affect the scaling asymptotically. 

 An approximate form for the moments are derived  by assuming a simple form of the density profile. 
Two decoherence parameters are defined which vanish in a power law manner  at extreme values of the disorder parameter $\alpha$ where  the usual 
DTQW is recovered. The above observations suggest the existence of a continuous phase transition taking place at $\alpha =0$ or $1$. 

Another interesting point to note is the universality of the scaling behaviour, the scaling exponent $3/2$ for $\la x^2 \ra$ and $1/2$ for $\la x \ra$ are  independent of $\alpha$, the parameter responsible for the localisation. This is in contrast to previous works where the scaling behaviour  was found to depend on the disorder parameter. 

 Randomness is expected to localise the system, however, the simultaneous 
lengthening of the steps could have played a counter-role. 
As we note from the case without randomness, making the step lengths 
larger does affect the behaviour of $f(x,t)$ substantially, however, the scaling behaviour remains same. Thus one can conclude that 
randomness is  the  key ingredient responsible for the novel scaling behaviour.  

While the slowing down of the walker  may not be  surprising  in  presence of 
the randomness, 
 a number of other results obtained in the study are not that obvious. 
The form of the distribution, especially the non-Gaussian behaviour close
to the origin, is not entirely predictable.
 The exponent $1.5$ for $\langle x^2\rangle $ similarly cannot be straight forwardly guessed.   The  universality of the exoponent value for any $\alpha \neq 0,1$ 
is also an interesting observation.  
Although the scaling exponent is not dependent on $\alpha$, its effect  is  present 
in the decoherence parameters $b_2$ and $b_4$.  The results for $\langle x^2 \rangle$ for different $\alpha$ as well as the behaviour of $b_4$   indicate the walker is maximally localised at $\alpha \simeq 0.8$.

A more general walk with  two independent parameters where a wider  spectrum of step lengths is  
considered also shows very similar results, a detailed study is in progress.  

In the present paper, the emphasis is on the  behaviour of the probability  distribution $f(x,t)$ and its moments. 
 The results  already show significant  departure from the conventional classical random and quantum walks. We believe these    studies
will inspire further characterization of the walk by addressing  
the issue  of  `quantumness' of the walk 
 \cite{Srikanth} using other measurements.

\medskip
Acknowledgement:
The author thanks Charles Bennett for a very important  suggestion.
Financial support from SERB project, Government of India,  is acknowledged.

\end{document}